\documentclass[referee]{raa}            

\usepackage{graphicx} 
\usepackage{natbib}
\usepackage{amssymb, amsmath}
\bibpunct{(}{)}{;}{a}{}{,}

\usepackage[pagebackref=true]{hyperref}

\begin{document}

\title{Results from Hubble parameter data: oscillating dark energy?}

\volnopage{Vol.0 (20xx) No.0, 000--000}      
\setcounter{page}{1}                         

\author{
    Rong-Jia Yang 
    \inst{1}
}

\institute{
    College of Physics Science and Technology, Hebei University, Baoding 071002, China; {\it yangrongjia@tsinghua.org.cn}
}

\titlerunning{Results from Hubble parameter data: oscillating dark energy?}

\abstract{
Using a model-independent analysis method which bases on the Lagrange mean value theorem for obtaining the derivative of the Hubble function, we analyze $H(z)$ parameter data with some restrictive conditions. We find that: (a) the Universe may experience an accelerated expansion with a confidence level greater than 5 $\sigma$ at redshift $z_{101}\in (0, 0.36)$ and greater than 1.9 $\sigma$ at redshifts $z_{3835}\in (1.3, 1.53)$ and $z_{3836}\in (1.43, 1.53)$, where $z_j<z_{ij}<z_i$ and $i$ marks the $i$-th Hubble parameter data we consider; (b) the Universe may experience a decelerated expansion with a confidence level greater than 1.5 $\sigma$ at redshift $z_{2012}\in (0.40, 0.52)$; (c) $w_{\rm{x}}\leq w_{\rm{t}}<-1$ with confidence level great than 1.6 $\sigma$ at redshift $z_{3836}\in (1.43, 1.53)$. These results indicate that the evolution of dark energy may be oscillatory.
}
\maketitle 



%
%
\section{Introduction}

Numerous independent cosmological observations have confirmed that the Universe is experiencing an accelerated
expansion (\citealt{Riess+etal+1998, Perlmutter:1998np, Tegmark:2003ud, WMAP:2012nax, Planck:2018vyg}). An unknown energy component, dubbed as dark energy, usually has been introduced in the framework of general relativity to explain this phenomenon. Vacuum energy is the simplest and most theoretically sound scenario of dark energy with an equation of state (EoS) $w_{\rm x}=p_{\rm x}/\rho_{\rm x}=-1$. If adding in cold dark matter, this model ($\Lambda$CDM) is consistent with the current astronomical observations, however, it suffers from the cosmological constant problem (\citealt{Carroll:2000fy}) and may age problem (\citealt{Yang:2009ae}) as well. Recently, Hubble tension (\citealt{Riess2019cxk}) suggests that $\Lambda$CDM may face new puzzles.

In the analysis of observational data, statistical methods, such as the maximum likelihood (\citealt{Yang:2009ae, Yang:2008hda, Yang:2013skk, Nesseris:2005ur, Lazkoz:2005sp}), are generally used to fit the model parameters. With statistical methods, we can obtain the best-fit values of model parameters. However, it is easy to count some interesting (possibly important) data off. In \cite{Yang:2023qsz}, a model-independent method without making assumptions about the EoS of dark energy or the Hubble function by using the Lagrange mean value theorem to obtain the derivative of the Hubble function was proposed to analyze $H(z)$ parameter data. When getting the deceleration parameter, a mid-value approximate method was adopted, but the errors caused by the method were also considered approximatively, which may have impacts on the results. Here, we further improve the model-independent analysis method proposed in \cite{Yang:2023qsz} and consider the errors caused by mid-value approximate method accurately to obtain the deceleration parameter. We find that the Universe may experience an accelerated expansion at higher redshifts ($1.3<z<1.53$), confirming the results obtained in \cite{Yang:2023qsz}.

The paper is organized as follows. In the next Section, we will present $H(z)$ parameter data and derive the equations needed to analyze these data. In Sec. III, We will provide the data and results obtained from the analysis. Finally, we will briefly summarize and discuss our results in Sec. IV.

\section{$H(z)$ parameter data and methodology}
In this Section, we will present 43 $H(z)$ parameter data obtained recently and improve the model-independent analysis method proposed in \cite{Yang:2023qsz} which is needed in the process of analyzing $H(z)$ parameter data, so as to try to explore the nature of dark energy.

\subsection{$H(z)$ parameter data}
$H(z)$ parameter data are widely used to constrain the parameters of dark energy models, see for example: \cite{Koussour:2024glo,Wei2019uhh,Qi:2023oxv,Li:2022cbk,Koussour:2024sio,He:2024jku,Li:2014yza,Yin:2018mvu,Figueroa:2008py,AlMamon:2018uby,Goswami:2023knh}. The data set we use consists of 1 $H_0$ measurement from supernova Ia (SNIa) observation, whose error is smaller than that used in \cite{Yang:2023qsz}, 27 $H(z)$ measurements inferred from the baryon acoustic oscillation (BAO) peak in the galaxy power spectrum, and 15 $H(z)$ measurements obtained by calculating the differential ages of galaxies, which is called cosmic chronometer. In three cases, the datasets are given with their $1$ $\sigma$ confidence interval, as listed in Table \ref{Hz}.
\begin{table}
\centering
\caption{Hubble parameter data from SN Ia observations, cosmic chronometers (DA), and BAO surveys (Clustering).}
\label{Hz}
\resizebox{\textwidth}{!}{
\begin{tabular}{llllll llllll}
\hline
index & $z$ & $H_{\rm{o}}(z)$[km s$^{-1}$Mpc$^{-1}$] & $\sigma_{\rm{H}}$ & Reference & Method & index & $z$ & $H_{\rm{o}}(z)$ & $\sigma_{\rm{H}}$ & Reference & Method \\
\hline
$z_1$ & 0 & 73.29 & 0.09 & \cite{Murakami:2023xuy} & SN Ia/Cepheid     & $z_{23}$ & 0.59 & 98.48  & 3.19 & \cite{Wang:2016wjr}& Clustering  \\
\hline
$z_2$ & 0.17 & 83 & 8 & \cite{Stern:2009ep} & DA                       & $z_{24}$ & 0.5929 & 104 & 13 & \cite{Moresco:2012jh} & DA \\
\hline
$z_3$ & 0.1797 & 75 & 4 & \cite{Moresco:2012jh} & DA                   & $z_{25}$ & 0.6 & 87.9 & 6.1 & \cite{Blake:2012pj} & Clustering \\
\hline
$z_4$ & 0.1993 & 75 & 5 & \cite{Moresco:2012jh} & DA                   & $z_{26}$ & 0.61 & 97.3  & 2.1 & \cite{Alam:2016hwk}& Clustering \\
\hline
$z_5$ & 0.24 & 79.69 & 2.65 & \cite{Gaztanaga:2008xz} & Clustering     & $z_{27}$ & 0.64 & 98.82  & 2.99 & \cite{Wang:2016wjr}& Clustering\\
\hline
$z_{6}$ & 0.3 & 81.7 & 6.22 & \cite{Oka:2013cba}& Clustering          & $z_{28}$ & 0.6797 & 92 & 8 & \cite{Moresco:2012jh} & DA \\
\hline
$z_{7}$ & 0.31 & 78.17 & 4.74 & \cite{Wang:2016wjr}& Clustering       & $z_{29}$ & 0.73 & 97.3 & 7 & \cite{Blake:2012pj} & Clustering \\
\hline
$z_{8}$ & 0.34 & 83.8 & 3.66 & \cite{Gaztanaga:2008xz} & Clustering   & $z_{30}$ & 0.7812 & 105 & 12 & \cite{Moresco:2012jh} & DA\\
\hline
$z_{9}$ & 0.35 & 82.7 & 8.4 & \cite{Chuang:2012qt} & Clustering       & $z_{31}$ & 0.8754 & 125 & 17 & \cite{Moresco:2012jh} & DA \\
\hline
$z_{10}$ & 0.36 & 79.93  & 3.39 & \cite{Wang:2016wjr}& Clustering      & $z_{32}$ & 0.978 & 113.72 & 14.63 & \cite{Zhao:2018gvb} & Clustering \\
\hline
$z_{11}$ & 0.38 & 81.5  & 1.9 & \cite{Alam:2016hwk}& Clustering        & $z_{33}$ & 1.037 & 154 & 20 & \cite{Moresco:2012jh} & DA\\
\hline
$z_{12}$ & 0.40 & 82.04 & 2.03 & \cite{Wang:2016wjr} & DA              & $z_{34}$ & 1.23 & 131.44 & 12.42 & \cite{Zhao:2018gvb} & Clustering \\
\hline
$z_{13}$ & 0.4293 & 91.8 & 5.3 & \cite{Moresco:2016mzx} & DA           & $z_{35}$ & 1.3 & 168 & 17 & \cite{Stern:2009ep} & DA \\
\hline
$z_{14}$ & 0.43 & 86.45 & 3.68 & \cite{Gaztanaga:2008xz} & Clustering  & $z_{36}$ & 1.43 & 177 & 18 & \cite{Stern:2009ep} & DA\\
\hline
$z_{15}$ & 0.44 & 82.6 & 7.8 & \cite{Blake:2012pj} & Clustering        & $z_{37}$ & 1.526 & 148.11 & 12.71 & \cite{Zhao:2018gvb} & Clustering \\
\hline
$z_{16}$ & 0.44 & 84.81 & 1.83 & \cite{Wang:2016wjr} & Clustering      & $z_{38}$ & 1.53 & 140 & 14 & \cite{Stern:2009ep} & DA \\
\hline
$z_{17}$ & 0.4783 & 80.9 & 9 & \cite{Moresco:2016mzx} &DA              &  $z_{39}$ & 1.944 & 172.63 & 14.79 & \cite{Zhao:2018gvb} & Clustering  \\
\hline
$z_{18}$ & 0.48 & 87.79 & 2.03 & \cite{Wang:2016wjr} & DA              &  $z_{40}$ & 2.3 & 224 & 8 & \cite{Busca:2012bu} & Clustering \\
\hline
$z_{19}$ & 0.51 & 90.4  & 1.9 & \cite{Alam:2016hwk}& Clustering        & $z_{41}$ & 2.33 & 224 & 8 & \cite{Bautista:2017zgn} & Clustering \\
\hline
$z_{20}$ & 0.52 & 94.35  & 2.65 & \cite{Wang:2016wjr}& Clustering      & $z_{42}$ & 2.34 & 222 & 7 & \cite{Delubac:2014aqe}& Clustering\\
\hline
$z_{21}$ & 0.56 & 93.33  & 2.32 & \cite{Wang:2016wjr}& Clustering      & $z_{43}$ & 2.36 & 226 & 8 & \cite{Font-Ribera:2013wce}& Clustering\\
\hline
$z_{22}$ &0.57 & 92.9 & 7.8 & \cite{Anderson:2013oza} & Clustering     & & & & & &                                                          \\
\hline
\hline
\end{tabular}
}
\end{table}

\subsection{Methodology.}
According to the Planck 2018 results, the spacetime is spatially flat: $\Omega_{\rm K0}=0.001\pm 0.002$ (\citealt{Planck:2018vyg}). Here we consider a spatially flat Friedmann-Robertson-Walker-Lema\^{i}tre (FRWL) metric
\begin{eqnarray}
\label{frwmet}
ds^2=-dt^2+a^2(t)\left[dr^2+r^2(d\theta^2+\sin^2\theta
d\phi^2)\right],
\end{eqnarray}
where $a(t)$ is the scale factor and the unit $c=1$ is used. The Friedmann equations take the form
\begin{eqnarray}
&&H^2=\frac{8\pi G}{3}\rho, \\
\label{acc}
&&\frac{\ddot{a}}{a}=-\frac{4\pi G}{3}\left(\rho+3p\right),
\end{eqnarray}
where the $H\equiv \dot{a}/a$ is the Hubble parameter and the dot denotes the derivative with respect to the cosmic time
$t$. The total energy density $\rho$ and pressure $p$ contain contributions coming from the nonrelativistic matter, radiation, and other
components. The Friedmann equations can be equivalently rewritten as
\begin{eqnarray}
\label{Frid}
\dot{H}=-4\pi G(\rho+p).
\end{eqnarray}
With $dz=-(1+z)Hdt$, we derive (\citealt{Yang:2023qsz})
\begin{eqnarray}
\label{6}
\frac{dH}{dz}=\frac{4\pi G(\rho+p)}{(1+z)H}=\frac{4\pi G\rho(1+w_{\rm{t}})}{(1+z)H},
\end{eqnarray}
where $w_{\rm{t}}$ is the total EoS. From this equation, one has $w_{\rm{x}}\leq w_{\rm{t}}\leq-1$ if $dH/dz\leq 0$, meaning that the Universe experiences an accelerated expansion. If $dH/dz>0$, however, we can't judge whether the Universe speeds up. At this moment, we need another physical quantity, the deceleration parameter, which is defined as
\begin{eqnarray}
\label{q}
q=-1+(1+z)\frac{1}{H}\frac{dH}{dz}.
\end{eqnarray}
Now if we have some Hubble parameter data $H_{\rm{o}}(z_i)$, a question naturally rise: how can we use them to directly determine $dH/dz$ or $q$? Supposing that $H(z)$ is the actual theoretical curve of the evolution of the Universe, we use a datum $H_{\rm{o}}(z_i)$ at redshift $z_i$ to approximate the value of $H(z_i)$ at 1 $\sigma_{{\rm{H}}_{i}}(z_i)$ confidence level.

Thinking of the Lagrange mean value theorem in Calculus, we have
\begin{eqnarray}
\label{h1}
H^{\prime}(z_{ij})\equiv\frac{dH}{dz}\big{|}_{z=z_{ij}}=\frac{H(z_i)-H(z_j)}{z_i-z_j},
\end{eqnarray}
where $z_j<z_{ij}<z_i$. Then we can approximate $H^{\prime}(z_{ij})$ as
\begin{eqnarray}
\label{ah}
H^{\prime}(z_{ij})\simeq \frac{H_{\rm{o}}(z_i)-H_{\rm{o}}(z_j)}{z_i-z_j},
\end{eqnarray}
at 1 $\sigma_{\rm{H}'}$ confidence level, where $\sigma_{\rm{H}'}$ is given by
\begin{eqnarray}
\label{eh}
\sigma_{\rm{H}^{\prime}}=\frac{\sqrt{\sigma^2_{\text{H}_{i}}+\sigma^2_{\text{H}_{j}}}}{z_i-z_j}.
\end{eqnarray}
Now considering the approximation of Eq. \eqref{q}
\begin{eqnarray}
\label{q}
q(z_{ij})=-1+\frac{1+z_{ij}}{H(z_{ij})}H^{\prime}(z_{ij})\simeq -1+\frac{1+z_{ij}}{H(z_{ij})}\frac{H_{\rm{o}}(z_i)-H_{\rm{o}}(z_j)}{z_i-z_j},
\end{eqnarray}
where $z_{ij}$ and $H(z_{ij})$ are unknown. Here we adopt mid-value approximate method proposed in \cite{Yang:2023qsz}: $z_{ij}\simeq (z_i+z_j)/2$ and $H(z_{ij})\simeq [H(z_i)+H(z_j)]/2\simeq [H_{\rm{o}}(z_i)+H_{\rm{o}}(z_j)]/2$, leading to
\begin{eqnarray}
\label{q1}
q(z_{ij})\simeq-1+\frac{\left(2+z_i+z_j\right)}{H_{\rm{o}}(z_i)+H_{\rm{o}}(z_j)}\frac{H_{\rm{o}}(z_i)-H_{\rm{o}}(z_j)}{z_i-z_j},
\end{eqnarray}
at 1 $\sigma_{\rm{q}}$ confidence level, where $\sigma_{\rm{q}}$ is given by
\begin{eqnarray}
\label{eq}
\sigma_{\text{q}}=\frac{2(2+z_i+z_j)}{[H_{\text{o}}(z_i)+H_{\text{o}}(z_j)]^2}\frac{\sqrt{H^2_{\text{o}}(z_j)\sigma^2_{\text{H}_{i}}+H^2_{\text{o}}(z_i)\sigma^2_{\text{H}_{j}}}}{z_i-z_j}.
\end{eqnarray}
This error formula is slightly difference from the Eq. (\ref{eq}) in \cite{Yang:2023qsz} where an approximation $H_{\text{o}}(z_i)\simeq H_{\text{o}}(z_j)$ was taken for the sake of simplicity in calculations. Similar approximate methods, like the mid-value approximation, were used in the literatures (\citealt{Yang:2020bpv,Li:2015nta}). To explain the credibility of this method, we take $\Lambda$CDM as an example: (1) taking $\Omega_{\rm m}=0.3$, $H_0=73$ km s$^{-1}$Mpc$^{-1}$, and $z_{i}=0.1$, and $z_{j}=0.6$, we find $[H(z_i)+H(z_j)]/2\simeq 88.96$ km s$^{-1}$Mpc$^{-1}$ and $H((z_i+z_j)/2)\simeq 87.54$ km s$^{-1}$Mpc$^{-1}$; (2) taking $z_{i}=1.8$, and $z_{j}=2.36$, we find $[H(z_i)+H(z_j)]/2\simeq 225.38$ km s$^{-1}$Mpc$^{-1}$ and $H((z_i+z_j)/2)\simeq 224.59$ km s$^{-1}$Mpc$^{-1}$. We see that the differences between those corresponding two values is much smaller than either of them: $\Delta H=[H(z_i)+H(z_j)]/2-H((z_i+z_j)/2)\ll H(z_j)<H(z_i)$.

Other methods, such as weighted average method (\citealt{Wei:2019uhh,Zheng:2019trp}) and Bayesian non-parametric method (\citealt{Shafieloo:2005nd,Shafieloo:2007cs,Shafieloo:2009hi,Busti:2014dua}), were also used to analyze Hubble parameter data. If there is summation or averaging in these methods, the errors will accumulate. The errors will not be accumulated when using mid-value approximate method, but they would be enlarged in general if $z_i-z_j$ is large. However, if the difference between $z_i$ and $z_j$ and the difference between $H_{\rm{o}}(z_i)$ and $H_{\rm{o}}(z_j)$ are reasonable, this approximate method in general is credible.

\begin{table}
\centering
\caption{$H^{\prime}(z)$ and $q(z)$ data obtained from $H(z)$ parameter data, where $z_{\rm m}=(z_i+z_j)/2$.}
\label{hq2}
\resizebox{\textwidth}{!}{
\begin{tabular}{llllllllllll}
\hline
index  & $z_{\rm m}$ & $H^{\prime}(z)$ & $\sigma_{\text{H}'}$ & $q(z)$ & $\sigma_{\text{q}}$ & index & $z_{\rm m}$  & $H^{\prime}(z)$ & $\sigma_{\text{H}'}$ & $q(z)$ & $\sigma_{\text{q}}$\\
\hline

$z_{31}\in (0, 0.1797)$ &0.0899 & 9.52 & 22.26 & -0.86 & 0.32 & $z_{125}\in (0.24, 0.40)$ & 0.32 &  14.69 & 20.86 & -0.76 & 0.34 \\
\hline
$z_{41}\in (0, 0.1993)$ &0.0997 & 8.58 & 25.09 & -0.87 & 0.37& $z_{145}\in (0.24, 0.43)$ & 0.335 & 35.58 & 23.87 & -0.43 & 0.38  \\
\hline
$z_{51}\in (0, 0.24)$ & 0.12 & 26.67 & 11.05 & -0.61 & 0.16 & $z_{155}\in (0.24, 0.44)$ & 0.34 &  14.55& 41.19 & -0.76 & 0.67  \\
\hline
$z_{61}\in (0, 0.3)$ & 0.15 & 28.03 & 20.74 & -0.58 & 0.29 &  $z_{165}\in (0.24, 0.44)$ & 0.34 &  25.6 & 16.1 & -0.58 & 0.27 \\
\hline
$z_{71}\in (0, 0.31)$ & 0.155& 15.74 & 15.29 & -0.76 & 0.23 &$z_{175}\in (0.24, 0.4783)$ & 0.3592 &  5.08 & 39.37 & -0.91 & 0.66 \\
\hline
$z_{81}\in (0, 0.34)$ & 0.17 & 30.91 & 10.77 & -0.54 & 0.15  & $z_{185}\in (0.24, 0.48)$ & 0.36 & 33.75 & 13.91 & -0.45 & 0.23 \\
\hline
$z_{91}\in (0, 0.35)$ & 0.175 & 26.89 & 24 & -0.59 & 0.34 &$z_{195}\in (0.24, 0.51)$ & 0.375 & 39.67 & 12.08 & -0.36 & 0.20 \\
\hline
$z_{101}\in (0, 0.36)$ & 0.18& 18.44 & 9.42 & -0.72 & 0.14 &$z_{215}\in (0.24, 0.56)$ & 0.4 & 42.625 & 11.01 & -0.31 & 0.18 \\
\hline
$z_{111}\in (0, 0.38)$ & 0.19& 21.61 & 5.01 & -0.67 & 0.07 &$z_{225}\in (0.24, 0.57)$ & 0.405 & 40.03 & 24.96 & -0.35 & 0.38 \\
\hline
$z_{121}\in (0, 0.4)$ & 0.2 & 21.88 & 5.08 & -0.66 & 0.07 & $z_{255}\in (0.24, 0.6)$ & 0.42 & 22.81 & 18.47 & -0.61 & 0.3 \\
\hline
$z_{131}\in (0.0, 0.4293)$ &0.2147& 43.12 & 12.35 & -0.37 & 0.16 &  $z_{265}\in (0.24, 0.61)$& 0.425 & 47.59 & 9.14 & -0.23 & 0.11\\
\hline
$z_{141}\in (0.0, 0.43)$ & 0.215& 30.6 & 8.56 & -0.53 & 0.12 &$z_{275}\in (0.24, 0.64)$ & 0.44 & 47.83 & 9.99 & -0.23 & 0.16  \\
\hline
$z_{151}\in (0.0, 0.44)$ &0.22& 21.16 & 17.73 & -0.67 & 0.26 & $z_{285}\in (0.24, 0.6797)$ & 0.4599 & 28 & 19.17 & -0.52 & 0.31 \\
\hline
$z_{161}\in (0.0, 0.44)$ & 0.22& 26.18 & 4.16 & -0.6 & 0.06 &$z_{295}\in (0.24, 0.73)$ & 0.485& 35.94 & 15.28 & -0.4 & 0.24 \\
\hline
$z_{171}\in (0.0, 0.4783)$ & 0.2392 & 15.91  & 18.82 & -0.74 & 0.29 &$z_{176}\in (0.3, 0.4783)$ & 0.3892 & -4.49 & 61.36 & -1.08 & 1.05\\
\hline
$z_{181}\in (0.0, 0.48)$ & 0.24& 30.21  & 4.23 & -0.53 & 0.06 & $z_{256}\in (0.3, 0.6)$ & 0.45 & 20.67 & 29.04 & -0.65 & 0.50 \\
\hline
$z_{191}\in (0.0, 0.51)$ &0.255& 33.55  & 3.73 & -0.49 & 0.05 & $z_{286}\in (0.3, 0.6797)$ & 0.4899 &  27.13 & 26.69 & -0.53 & 0.45 \\
\hline

$z_{201}\in (0.0, 0.52)$ &0.26& 40.5  & 5.1 & -0.39 & 0.07 & $z_{296}\in (0.3, 0.73)$ & 0.515 & 36.28 & 21.78 & -0.39 & 0.37 \\
\hline

$z_{72}\in (0.17, 0.31)$ &0.24 & -34.5 & 66.42 & -1.53 & 1 & $z_{168}\in (0.34, 0.44)$ & 0.39 & 10.1 & 40.92 & -0.83 & 0.68 \\
\hline
$z_{82}\in (0.17, 0.34)$ & 0.255 & 4.71 & 51.75 & -0.93 & 0.78 & $z_{178}\in (0.34, 0.4783)$ & 0.4092 & -20.97 & 70.25 & -1.36 & 1.22\\
\hline
$z_{92}\in (0.17, 0.35)$ & 0.26 & -1.67 & 64.44 & -1.03 & 0.98 & $z_{188}\in (0.34, 0.48)$ & 0.41 &  28.5 & 29.89 & -0.53 & 0.5 \\
\hline
$z_{102}\in (0.17, 0.36)$ & 0.265 &-16.16 & 45.73 & -1.25 & 0.7 &$z_{258}\in (0.34, 0.6)$ & 0.47 & 15.77 & 27.36 & -0.73 & 0.46\\
\hline
$z_{112}\in (0.17, 0.38)$ & 0.275 & -7.14 & 39.15 & -1.11 & 0.6  & $z_{288}\in (0.34, 0.6797)$ & 0.5099 & 24.14 & 25.9 & -0.59 & 0.43\\
\hline
$z_{122}\in (0.17, 0.40)$ & 0.285 & -4.17 & 35.88 & -1.06 & 0.56 & $z_{298}\in (0.34, 0.73)$ & 0.535 & 34.62 & 20.25 & -0.41 & 0.33 \\
\hline
$z_{142}\in (0.17, 0.43)$ & 0.3 & 13.26 & 33.87 & -0.8 & 0.53 & $z_{2010}\in (0.36, 0.52)$ & 0.44 & 90.13 & 26.89 & 0.49 & 0.45 \\
\hline
$z_{152}\in (0.17, 0.44)$ & 0.305& -1.48 & 41.38 & -1.02 & 0.65 & $z_{2011}\in (0.38, 0.52)$ & 0.45 & 91.79 & 23.29 & 0.51 & 0.38 \\
\hline
$z_{162}\in (0.17, 0.44)$ & 0.305& 6.7 & 30.39 & -0.9 & 0.48 & $z_{2311}\in (0.38, 0.59)$ & 0.485 & 80.86 & 17.68 & 0.33 & 0.28 \\
\hline
$z_{172}\in (0.17, 0.4783)$ & 0.3242 &-6.81 & 39.06 & -1.11 & 0.63 & $z_{2012}\in (0.40, 0.52)$ & 0.46&  102.58 & 27.82 & 0.7 & 0.45 \\
\hline
$z_{182}\in (0.17, 0.48)$ & 0.325& 15.45 & 26.62 & -0.76 & 0.42 & $z_{2312}\in (0.40, 0.59)$ & 0.495 & 86.53 & 19.9 & 0.43 & 0.32 \\
\hline
$z_{192}\in (0.17, 0.51)$ & 0.34 & 21.76 & 24.18 & -0.66 & 0.39 & $z_{2316}\in (0.44, 0.59)$ & 0.515 & 91.13 & 24.52 & 0.51 & 0.39 \\
\hline
$z_{202}\in (0.17, 0.52)$ & 0.345 & 32.43 & 24.08 & -0.51 & 0.38 & $z_{2318}\in (0.48, 0.59)$ & 0.535 & 97.18 & 34.37 & 0.6 & 0.55 \\
\hline
$z_{212}\in (0.17, 0.56)$& 0.365 & 26.49 & 21.36 & -0.59 & 0.35 & $z_{3321}\in (0.56, 1.037)$ & 0.7985 & 127.19 & 42.21 & 0.85 & 0.47\\
\hline
$z_{222}\in (0.17, 0.57)$& 0.37 & 24.75 & 27.93 & -0.61 & 0.44 & $z_{3323}\in (0.59, 1.037)$ & 0.8135 & 124.21 & 45.31 & 0.78 & 0.52 \\
\hline
$z_{232}\in (0.17, 0.59)$ & 0.38 & 36.86 & 20.51 & -0.44 & 0.33 &$z_{3326}\in (0.61, 1.037)$ & 0.8235 & 132.79 & 47.1 & 0.93 & 0.53 \\
\hline
$z_{252}\in (0.17, 0.6)$& 0.385 & 11.4 & 23.4 & -0.82 & 0.38 & $z_{3327}\in (0.64, 1.037)$ & 0.8385 & 138.99 & 50.94 & 1.02 & 0.59 \\
\hline
$z_{262}\in (0.17, 0.61)$ & 0.39 & 32.5 & 18.8 & -0.5 & 0.31 & $z_{3330}\in (0.7812, 1.037)$ & 0.9091 & 191.56 & 91.18 & 1.82 & 1.24 \\
\hline
$z_{272}\in (0.17, 0.64)$ &0.405 & 33.66 & 18.17 & -0.48 & 0.3 & $z_{3532}\in (0.978, 1.3)$ & 1.139 & 168.57 & 69.65 & 1.56 & 1.05 \\
\hline

$z_{103}\in (0.1797, 0.36)$ & 0.2699 & 27.34 & 29.08 & -0.55 & 0.48 & $z_{3632}\in (0.978, 1.43)$ & 1.204 & 140 & 51.32 & 1.12 & 0.76 \\
\hline
$z_{113}\in (0.1797, 0.38)$ & 0.2799 & 32.45 & 22.11 & -0.47 & 0.37 & $z_{3433}\in (1.037, 1.23)$ & 1.1335&  -116.89 & 121.98 & -2.75 & 1.76\\
\hline
$z_{123}\in (0.1797, 0.40)$ & 0.2899 & 31.96 & 20.36 & -0.48 & 0.34 & $z_{3733}\in (1.037, 1.526)$ & 1.2815 & -12.05 & 48.46 & -1.18 & 0.73 \\
\hline
$z_{163}\in (0.1797, 0.44)$ & 0.3099 & 37.69 & 16.9 & -0.38 & 0.29 & $z_{3833}\in (1.037, 1.53)$ & 1.2835 & -28.4 & 49.52 & -1.44 & 0.76 \\
\hline
$z_{173}\in (0.1797, 0.4783)$ & 0.329 & 19.76 & 32.98 & -0.66 & 0.55 &  $z_{3634}\in (1.23, 1.43)$ & 1.33 & 227.8 & 109.35 & 2.44 & 1.58 \\
\hline
$z_{183}\in (0.1797, 0.48)$ & 0.3299 & 42.59 & 14.94 & -0.3 & 0.26 & $z_{3735}\in (1.3, 1.526)$ & 1.413 & -88.01 & 93.92 & -2.34 & 1.41\\
\hline
$z_{193}\in (0.1797, 0.51)$ & 0.3449 & 46.62 & 13.41 & -0.24 & 0.23 & $z_{3835}\in (1.3, 1.53)$ & 1.415 & -121.74 & 95.75 & -2.91 & 1.48\\
\hline
$z_{253}\in (0.1797, 0.6)$ & 0.3899 & 30.69 & 17.36 & -0.48 & 0.29  & $z_{3736}\in (1.43, 1.526)$ & 1.478 & -300.94 & 229.53 & -5.59 & 3.41\\
\hline
$z_{283}\in (0.1797, 0.6797)$ & 0.4297 & 34 & 17.89 & -0.42 & 0.29 & $z_{3836}\in (1.43, 1.53)$ & 1.48 & -370 & 228.04 & -6.79 & 3.49\\
\hline

$z_{174}\in (0.1993, 0.4783)$ & 0.3388 & 21.15 & 36.9 & -0.64 & 0.62 & $z_{3934}\in (1.43, 1.944)$ & 1.687 & -8.52 & 45.32 & -1.13 & 0.70 \\
\hline
$z_{254}\in (0.1993, 0.6)$ & 0.3997 & 32.19 & 19.68 & -0.45 & 0.33 & $z_{4039}\in (1.944, 2.3)$ & 2.122 & 144.3 & 47.23 & 1.27 & 0.8\\
\hline
$z_{284}\in (0.1993, 0.6797)$ & 0.4395 & 35.39 & 19.64 & -0.39 & 0.32 &$z_{4139}\in (1.944, 2.33)$ & 2.137 & 133.08 & 43.56 & 1.11 & 0.74 \\
\hline

$z_{105}\in (0.24, 0.36)$ & 0.3 & 2 & 35.86 & -0.97 & 0.58 & $z_{4239}\in (1.944, 2.34)$ & 2.142 & 124.67 & 41.32 & 0.99 & 0.71 \\
\hline
$z_{115}\in (0.24, 0.38)$ & 0.31 &  12.93 & 23.29 & -0.79 & 0.38 &$z_{4339}\in (1.944, 2.36)$ & 2.152 & 128.29 & 40.42 & 1.03 & 0.69 \\
\hline
\end{tabular}
}
\end{table}

\section{Applications}
Now using Eqs. (\ref{h1}), (\ref{q1}), (\ref{eh}), and (\ref{eq}), we analyze the observational Hubble parameter data. According to Eq. (\ref{h1}), if $z_i-z_j\ll 1$, $H^{\prime}(z_{ij})$ would be large, implying that the systematic errors will be amplified. So in the process of analyzing the Hubble parameter data, we consider the following limitations to make the results credible: $0.1\lesssim z_i-z_j\lesssim 0.5$, $\sigma_{\rm{H}}\leq 10$ if $H\leq 100$, and $\sigma_{\rm{H}}\leq 20$ if $H\geq 100$. We do this based on the following considerations: firstly, the errors of the Hubble parameter data are relatively large; secondly, the Lagrange mean value theorem holds for any redshift interval, but when the redshift interval is relatively large, the error of Eq. (\ref{ah}) can be reduced. At the same time, we should also maintain reasonable differences of $z_i-z_j$ and $H_{\rm{o}}(z_i)-H_{\rm{o}}(z_j)$ to make the approximation of the Eq. (\ref{q1}) reliable. Compared to the limitations taken in \cite{Yang:2023qsz}, the scope of data analyzed here has been expanded. The obtained $H^{\prime}$ and $q$ data with 1 $\sigma$ confidence level are listed in Table \ref{hq2}. The results are consistent with those obtained in \cite{Yang:2023qsz}. However, comparing with the 78 data (39 $H^{\prime}(z)$ data and 39 $q(z)$ data) obtained in \cite{Yang:2023qsz}, here we present 204 data (102 $H^{\prime}(z)$ data and 102 $q(z)$ data) with considerations of the errors caused by mid-value approximate method, which makes the obtained data more reliable. From these data, we emphasize the following results:

(a) The Universe may experience an accelerated expansion during the period $0< z< 0.36$: see for example, the expansion of the Universe may speed up with a confidence level greater than 3.8 $\sigma$ at redshift $z_{51}\in (0, 0.24)$; greater than 3.6 $\sigma$ at $z_{81}\in (0, 0.34)$; greater than 5 $\sigma$ at redshifts $z_{101}\in (0, 0.36)$ and $z_{201}\in (0, 0.52)$.

(b) The Universe may experience a decelerated expansion during the period $0.36< z< 1.037$: see for example, the expansion of the Universe may speed down with a confidence level greater than 1.5 $\sigma$ at redshift $z_{2012}\in (0.40, 0.52)$ (greater than the confidence level obtained from the corresponding data $z_{3420}$ in \cite{Yang:2023qsz}; greater than 1.8 $\sigma$ at redshift $z_{3321}\in (0.56, 1.037)$; greater than 1.7 $\sigma$ at redshifts $z_{3326}\in (0.61, 1.037)$ and $z_{3327}\in (0.64, 1.037)$.

(c) The Universe may experience an accelerated expansion during the period $1.037< z< 1.944$: see for example, the expansion of the Universe may speed up with a confidence level greater than 1.5 $\sigma$ at redshift $z_{3433}\in (1.037, 1.23)$; greater than 1.6 $\sigma$ at redshifts $z_{3733}\in (1.037, 1.526)$, $z_{3735}\in (1.3, 1.526)$, $z_{3736}\in (1.43, 1.526)$, and $z_{3934}\in (1.43, 1.944)$; greater than 1.8 $\sigma$ at redshift $z_{3833}\in (1.037, 1.53)$; greater than 1.9 $\sigma$ at redshifts $z_{3835}\in (1.3, 1.53)$ and $z_{3836}\in (1.43, 1.53)$. These confidence levels are slightly lower than those obtained in \cite{Yang:2023qsz}.

(d) Since $H^{\prime}(z)< 0$, we find $w_{\rm{x}}\leq w_{\rm{t}}< -1$ with a confidence level greater than 1.2 $\sigma$ at redshifts $z_{3835}\in(1.3, 1.53)$ and $z_{3736}\in(1.43, 1.526)$. However, we infer that $w_{\rm{x}}\leq w_{\rm{t}}<-1$ with confidence level great than 1.6 $\sigma$ at redshift $z_{3836}\in (1.43, 1.53)$. These confidence levels are the same obtained in \cite{Yang:2023qsz}.

Results (a), (b), (c), and (d) suggest that the behavior of dark energy may be oscillatory. Based on DESI DR2 data, \cite{2025NatAs.tmp..195G,2025arXiv250615739Y} also obtained similar results. And \cite{2025arXiv250621010N} theoretically realizes the $f(R)$ gravity with oscillatory behavior.

\section{Conclusions and discussions}
Using a model-independent analysis method improved from that proposed in \cite{Yang:2023qsz} and taking into account the errors caused by mid-value approximation to increase the credibility of data, we have analyzed $H(z)$ parameter data with some restrictive conditions. Comparing with the 78 data obtained in \cite{Yang:2023qsz}, we have presented 204 data here. In order to make our conclusion more credible, we have provided the corresponding confidence levels which had not been calculated out in \cite{Yang:2023qsz}.

From obtained data, we have found that: (a) the Universe may experience an accelerated expansion during the period $0< z< 0.36$; (b) the Universe may experience a decelerated expansion during the period $0.36< z< 1.037$; (c) the Universe may experience an accelerated expansion during the period $1.037< z< 1.944$; (d) The EoS of dark energy may be less than $-1$ during the period $1.3<z<1.53$. Other studies have also suggested a dynamical dark energy, see for example: \cite{Yang:2023qsz,Wei:2006ut,Wei:2007ws,Zhang:2007uha,Yang:2024kdo,Zhao:2017cud,DESI:2024aqx,DESI:2024mwx,Feng:2004ad,Zhang:2023neo,2025JHEAp..4700398O}. Unlike these results, our studies suggest that the behavior of dark energy may be oscillatory, which is consistent with the research findings in \cite{2025NatAs.tmp..195G,2025arXiv250615739Y} based on DESI DR2 data.

The $q(z)$ data and some $H^{\prime}$ data obtained here could be used to constrain cosmological models. The reliability of the results obtained here depends on the $H(z)$ parameter data. More and more accurate $H(z)$ data are needed to validate our results in future researches. The data and the method presented here could be helpful in exploring the nature of dark energy.

\section*{Acknowledgments}
This study is supported in part by National Natural Science Foundation of China (Grant No. 12333008) and Hebei Provincial Natural Science Foundation of China (Grant No. A2021201034).

\bibliographystyle{raa}
\bibliography{H1}

\label{lastpage}

\end{document}